\documentclass[aps,prl,twocolumn,floats,floatfix,superscriptaddress,showpacs]{revtex4}
\usepackage{amsmath}
\usepackage{bm}
\usepackage{dcolumn}
\usepackage{psfig}
\usepackage{graphicx}

\newcommand{\etal}{{\it et al.\/}}
\renewcommand{\Re}{{\rm{Re}}}
\renewcommand{\Im}{{\rm{Im}}}

\begin{document}
\vspace{1.0cm}

\title{Recoil Polarization for Delta Excitation in Pion Electroproduction}
\author{J.~J.~Kelly}
\affiliation{Department of Physics, University of Maryland, College Park, Maryland 20742, USA}

\author{R.~E.~Roch\'e}
\affiliation{Florida State University, Tallahassee, Florida 32306, USA}

\author{Z.~Chai}
\affiliation{Massachusetts Institute of Technology, Cambridge, Massachusetts 02139, USA}

\author{M.~K.~Jones}
\affiliation{Thomas Jefferson National Accelerator Facility, Newport News, Virginia 23606, USA}

\author{O.~Gayou}
\affiliation{Massachusetts Institute of Technology, Cambridge, Massachusetts 02139, USA}

\author{A.~J.~Sarty}
\affiliation{Saint Mary's University, Halifax, Nova Scotia, Canada B3H 3C3}

\author{S.~Frullani}
\affiliation{Istituto Nazionale di Fisica Nucleare, Sezione Sanit\`a 
and Istituto Superiore di Sanit\`a, Physics Laboratory, 00161 Roma, Italy}

\author{K.~Aniol}
\affiliation{California State University Los Angeles, Los Angeles, California 90032, USA}

\author{E.~J.~Beise}
\affiliation{Department of Physics, University of Maryland, College Park, Maryland 20742, USA}

\author{F.~Benmokhtar}
\affiliation{Rutgers, The State University of New Jersey, Piscataway, New Jersey 08854, USA}

\author{W.~Bertozzi}
\affiliation{Massachusetts Institute of Technology, Cambridge, Massachusetts 02139, USA}

\author{W.~U.~Boeglin}
\affiliation{Florida International University, Miami, Florida 33199, USA}

\author{T.~Botto}
\affiliation{University of Athens, Athens, Greece}

\author{E.~J.~Brash}
\affiliation{University of Regina, Regina, Saskatchewan, Canada S4S 0A2}

\author{H.~Breuer}
\affiliation{Department of Physics, University of Maryland, College Park, Maryland 20742, USA}

\author{E.~Brown}
\affiliation{University of Georgia, Athens, Georgia 30602, USA}

\author{E.~Burtin}
\affiliation{CEA Saclay, F-91191 Gif-sur-Yvette, France}

\author{J.~R.~Calarco}
\affiliation{University of New Hampshire, Durham, New Hampshire 03824, USA}

\author{C.~Cavata}
\affiliation{CEA Saclay, F-91191 Gif-sur-Yvette, France}

\author{C.~C.~Chang}
\affiliation{Department of Physics, University of Maryland, College Park, Maryland 20742, USA}

\author{N.~S.~Chant}
\affiliation{Department of Physics, University of Maryland, College Park, Maryland 20742, USA}

\author{J.-P.~Chen}
\affiliation{Thomas Jefferson National Accelerator Facility, Newport News, Virginia 23606, USA}

\author{M.~Coman}
\affiliation{Florida International University, Miami, Florida 33199, USA}

\author{D.~Crovelli}
\affiliation{Rutgers, The State University of New Jersey, Piscataway, New Jersey 08854, USA}

\author{R.~De Leo}
\affiliation{Istituto Nazionale di Fisica Nucleare, Sezione Sanit\`a 
and Istituto Superiore di Sanit\`a, Physics Laboratory, 00161 Roma, Italy}

\author{S.~Dieterich}
\affiliation{Rutgers, The State University of New Jersey, Piscataway, New Jersey 08854, USA}

\author{S.~Escoffier}
\affiliation{CEA Saclay, F-91191 Gif-sur-Yvette, France}

\author{K.~G.~Fissum}
\affiliation{University of Lund, Box 118, SE-221 00 Lund, Sweden}

\author{V.~Garde}
\affiliation{Universit\'e Blaise Pascal Clermont Ferrand et CNRS/IN2P3 LPC 63, 177 Aubi\`ere Cedex, France}

\author{F.~Garibaldi}
\affiliation{Istituto Nazionale di Fisica Nucleare, Sezione Sanit\`a 
and Istituto Superiore di Sanit\`a, Physics Laboratory, 00161 Roma, Italy}

\author{S.~Georgakopoulos}
\affiliation{University of Athens, Athens, Greece}

\author{S.~Gilad}
\affiliation{Massachusetts Institute of Technology, Cambridge, Massachusetts 02139, USA}

\author{R.~Gilman}
\affiliation{Rutgers, The State University of New Jersey, Piscataway, New Jersey 08854, USA}

\author{C.~Glashausser}
\affiliation{Rutgers, The State University of New Jersey, Piscataway, New Jersey 08854, USA}

\author{J.-O.~Hansen}
\affiliation{Thomas Jefferson National Accelerator Facility, Newport News, Virginia 23606, USA}

\author{D.~W.~Higinbotham}
\affiliation{Massachusetts Institute of Technology, Cambridge, Massachusetts 02139, USA}

\author{A.~Hotta}
\affiliation{University of Massachusetts, Amherst, Massachusetts 01003, USA}

\author{G.~M.~Huber}
\affiliation{University of Regina, Regina, Saskatchewan, Canada S4S 0A2}

\author{H.~Ibrahim}
\affiliation{Old Dominion University, Norfolk, Virginia 23529, USA}

\author{M.~Iodice}
\affiliation{Istituto Nazionale di Fisica Nucleare, Sezione Sanit\`a 
and Istituto Superiore di Sanit\`a, Physics Laboratory, 00161 Roma, Italy}

\author{C.~W.~de~Jager}
\affiliation{Thomas Jefferson National Accelerator Facility, Newport News, Virginia 23606, USA}

\author{X.~Jiang}
\affiliation{Rutgers, The State University of New Jersey, Piscataway, New Jersey 08854, USA}

\author{A.~Klimenko}
\affiliation{Old Dominion University, Norfolk, Virginia 23529, USA}

\author{S.~Kozlov}
\affiliation{University of Regina, Regina, Saskatchewan, Canada S4S 0A2}

\author{G.~Kumbartzki}
\affiliation{Rutgers, The State University of New Jersey, Piscataway, New Jersey 08854, USA}

\author{M.~Kuss}
\affiliation{Thomas Jefferson National Accelerator Facility, Newport News, Virginia 23606, USA}

\author{L.~Lagamba}
\affiliation{Istituto Nazionale di Fisica Nucleare, Sezione Sanit\`a 
and Istituto Superiore di Sanit\`a, Physics Laboratory, 00161 Roma, Italy}

\author{G.~Laveissi\`ere}
\affiliation{Universit\'e Blaise Pascal Clermont Ferrand et CNRS/IN2P3 LPC 63, 177 Aubi\`ere Cedex, France}

\author{J.~J.~LeRose}
\affiliation{Thomas Jefferson National Accelerator Facility, Newport News, Virginia 23606, USA}

\author{R.~A.~Lindgren}
\affiliation{University of Virginia, Charlottesville, Virginia 22901, USA}

\author{N.~Liyanage}
\affiliation{Thomas Jefferson National Accelerator Facility, Newport News, Virginia 23606, USA}

\author{G.~J.~Lolos}
\affiliation{University of Regina, Regina, Saskatchewan, Canada S4S 0A2}

\author{R.~W.~Lourie}
\affiliation{Renaissance Technologies Corporation, Setauket, New York 11733, USA}

\author{D.~J.~Margaziotis}
\affiliation{California State University Los Angeles, Los Angeles, California 90032, USA}

\author{F.~Marie}
\affiliation{CEA Saclay, F-91191 Gif-sur-Yvette, France}

\author{P.~Markowitz}
\affiliation{Florida International University, Miami, Florida 33199, USA}

\author{S.~McAleer}
\affiliation{Florida State University, Tallahassee, Florida 32306, USA}

\author{D.~Meekins}
\affiliation{Florida State University, Tallahassee, Florida 32306, USA}

\author{R.~Michaels}
\affiliation{Thomas Jefferson National Accelerator Facility, Newport News, Virginia 23606, USA}

\author{B.~D.~Milbrath}
\affiliation{Eastern Kentucky University, Richmond, Kentucky 40475, USA}

\author{J.~Mitchell}
\affiliation{Thomas Jefferson National Accelerator Facility, Newport News, Virginia 23606, USA}

\author{J.~Nappa}
\affiliation{Rutgers, The State University of New Jersey, Piscataway, New Jersey 08854, USA}

\author{D.~Neyret}
\affiliation{CEA Saclay, F-91191 Gif-sur-Yvette, France}

\author{C.~F.~Perdrisat}
\affiliation{College of William and Mary, Williamsburg, Virginia 23187, USA}

\author{M.~Potokar}
\affiliation{University of Ljubljana, Kongresni trg 12, SI-1000 Ljubljana, Slovenia}

\author{V.~A.~Punjabi}
\affiliation{Norfolk State University, Norfolk, Virginia 23504, USA}

\author{T.~Pussieux}
\affiliation{CEA Saclay, F-91191 Gif-sur-Yvette, France}

\author{R.~D.~Ransome}
\affiliation{Rutgers, The State University of New Jersey, Piscataway, New Jersey 08854, USA}

\author{P.~G.~Roos}
\affiliation{Department of Physics, University of Maryland, College Park, Maryland 20742, USA}

\author{M.~Rvachev}
\affiliation{Massachusetts Institute of Technology, Cambridge, Massachusetts 02139, USA}

\author{A.~Saha}
\affiliation{Thomas Jefferson National Accelerator Facility, Newport News, Virginia 23606, USA}

\author{S.~\v{S}irca}
\affiliation{Massachusetts Institute of Technology, Cambridge, Massachusetts 02139, USA}

\author{R.~Suleiman}
\affiliation{Massachusetts Institute of Technology, Cambridge, Massachusetts 02139, USA}

\author{S.~Strauch}
\affiliation{Rutgers, The State University of New Jersey, Piscataway, New Jersey 08854, USA}

\author{J.~A.~Templon}
\affiliation{University of Georgia, Athens, Georgia 30602, USA}

\author{L.~Todor}
\affiliation{Old Dominion University, Norfolk, Virginia 23529, USA}

\author{P.~E.~Ulmer}
\affiliation{Old Dominion University, Norfolk, Virginia 23529, USA}

\author{G.~M.~Urciuoli}
\affiliation{Istituto Nazionale di Fisica Nucleare, Sezione Sanit\`a 
and Istituto Superiore di Sanit\`a, Physics Laboratory, 00161 Roma, Italy}

\author{L.~B.~Weinstein}
\affiliation{Old Dominion University, Norfolk, Virginia 23529, USA}

\author{K.~Wijesooriya}
\affiliation{University of Illinois at Urbana-Champaign, Urbana, Illinois 61801, USA}

\author{B.~Wojtsekhowski}
\affiliation{Thomas Jefferson National Accelerator Facility, Newport News, Virginia 23606, USA}

\author{X.~Zheng}
\affiliation{Massachusetts Institute of Technology, Cambridge, Massachusetts 02139, USA}

\author{L.~Zhu}
\affiliation{Massachusetts Institute of Technology, Cambridge, Massachusetts 02139, USA}

\collaboration{The Jefferson Laboratory E91011 and Hall A Collaborations}
\noaffiliation

\date{May 23, 2005}

\begin{abstract}
We measured angular distributions of recoil-polarization response functions
for neutral pion electroproduction for $W=1.23$ GeV at 
$Q^2 = 1.0$ (GeV/$c$)$^2$, 
obtaining 14 separated response functions plus 2 Rosenbluth combinations;
of these, 12 have been observed for the first time.
Dynamical models do not describe quantities governed by imaginary
parts of interference products well, indicating the need for adjusting
magnitudes and phases for nonresonant amplitudes.
We performed a nearly model-independent multipole analysis 
and obtained values for $\Re S_{1+}/M_{1+} = -(6.84 \pm 0.15)\%$ 
and $\Re E_{1+}/M_{1+} = -(2.91 \pm 0.19)\%$ that are distinctly
different from those from the traditional Legendre analysis based
upon $M_{1+}$ dominance and $\ell_\pi \leq 1$ truncation.
\end{abstract}
\pacs{14.20.Gk,13.60.Le,13.40.Gp,13.88.+e}

\maketitle

Insight into QCD-inspired models of hadron structure can be obtained by 
studying the properties of the nucleon and its low-lying excited states using
electromagnetic reactions with modest spacelike four-momentum transfer, $Q^2$.
In the very simplest models, quark-quark interactions with SU(6) spin-flavor 
symmetry suggest that the dominant configuration for the nucleon consists of 
three quarks in an $S$-state with orbital and total angular momenta $L=0$ 
and $J=1/2$, while the lowest excited state, 
the $\Delta$ resonance at $M_\Delta = 1.232$ GeV with $J=3/2$,
is reached by flipping the spin of a single quark and leaving $L=0$. 
Thus, the pion electroproduction reaction for invariant mass 
$W \approx M_\Delta$ is dominated by the $M_{1+}$ multipole amplitude.
However, the $M_\Delta - M_N$ mass splitting and the nonzero neutron electric 
form factor clearly demonstrate that SU(6) symmetry is broken by color
hyperfine interactions that introduce $D$-state admixtures with 
$L=2$ into these wave functions \cite{Isgur82}.
Although quadrupole configurations cannot be observed directly in
elastic electron scattering by the nucleon, their presence in both wave 
functions contributes to $S_{1+}$ and $E_{1+}$ multipole amplitudes for 
electroexcitation of the $\Delta$.
Additional contributions to these smaller multipoles may also arise 
from meson and gluon exchange currents between quarks \cite{Buchmann00}
or coupling to the pion cloud outside the quark core 
\cite{Fiolhais96,Kamalov99}.
Recently, it has also become possible to calculate $N$ to $\Delta$ transition
form factors using lattice QCD, 
albeit in quenched approximation \cite{Alexandrou05}.

The relative strength of the quadrupole amplitudes is normally quoted
in terms of the ratios $\text{SMR} = \Re S_{1+}/M_{1+}$ and 
$\text{EMR} = \Re E_{1+}/M_{1+}$ evaluated for isospin $3/2$ at $W=M_\Delta$,
but isospin analysis would require data for the $n\pi^+$ channel also.
Fortunately, model calculations show that the isospin $1/2$ contribution
to these ratios is almost negligible.
For example, one obtains $(\text{SMR},\text{EMR}) = (-6.71\%,-1.62\%)$ for 
isospin $3/2$ compared  with $(-6.73\%,-1.65\%)$ for the $p\pi^0$ channel 
using MAID2003 \cite{MAID2003} at $Q^2 = 1$ (GeV/$c$)$^2$.
Therefore, we quote results for the $p\pi^0$ channel without making
model-dependent corrections for the isospin $1/2$ contamination.

Most previous measurements of the quadrupole amplitudes for $\Delta$ 
electroexcitation fit Legendre coefficients to angular distributions 
of the unpolarized cross section for pion production and employ a
truncation that assumes: 1) only partial waves with $\ell_\pi \leq 1$
contribute and 2) terms not involving $M_{1+}$ can be omitted.  
However, a more detailed analysis using models shows that neither 
assumption is sufficiently accurate \cite{Kelly05d}.
Therefore, it is important to obtain data that are complete enough for 
nearly model-independent multipole analysis without relying upon $sp$ 
truncation or $M_{1+}$ dominance. 

More detailed information about the nonresonant background can be
obtained from polarization measurements that are sensitive to the
relative phase between resonant and nonresonant amplitudes.
This phase information is needed to test dynamical models that attempt
to distinguish between the intrinsic properties of a resonance and the
effects of rescattering.
A few previous measurements of recoil polarization have been made for 
low $Q^2$ with the proton parallel to the momentum transfer 
\cite{Warren98,Pospischil01}, but their kinematic coverage is quite limited.
Several recent measurements of beam analyzing power have also been made
\cite{Bartsch02,Joo03,Kunz03}.
Those experiments demonstrated that recent dynamical models do not 
describe the nonresonant background well.
More generally, there are 18 independent response functions for the
$p(\vec{e},e^\prime \vec{p})\pi^0$ reaction, of which half are sensitive
to real and half to imaginary parts of products of multipole amplitudes
\cite{Raskin89}.
In this Letter we report angular distributions for 14 separated
response functions plus 2 Rosenbluth combinations for $W = 1.23$ GeV
at $Q^2 = 1.0$ (GeV/$c$)$^2$ that are sufficiently complete to perform a 
phenomenological multipole analysis;
twelve of these response functions are obtained here for the first time.
Data for a wider range of $W$ will be presented later in a more detailed paper.

The observables for recoil polarization can be resolved into response 
functions according to
\begin{subequations}
\label{eq:obs}
\begin{eqnarray}
\bar{\sigma}    &=& \nu_0 \left[ \nu_L R_L + \nu_T R_T 
       + \nu_{LT}  R_{LT} \sin \theta \cos \phi \right. \\ \nonumber
    &+&   \left.  \nu_{TT} R_{TT} \sin^2 \theta \cos 2\phi \right]  \\
 A \bar{\sigma}  &=&  \nu_0 
\left[ \nu_{LT}^\prime  R_{LT}^\prime \sin \theta \sin\phi  \right] \\
 P_n \bar{\sigma} &=& \nu_0 \left[
   (\nu_L  R_L^n + \nu_T R_T^n) \sin\theta  \right. \\ \nonumber
  &+& \left. \nu_{LT} R_{LT}^n \cos\phi
        +\nu_{TT} R_{TT}^n \sin\theta\cos 2\phi  \right]  \\
 P_{\ell} \bar{\sigma} &=&  \nu_0 \left[ 
           \nu_{LT} R_{LT}^{\ell} \sin\theta \sin\phi \right. \\ \nonumber
         &+& \left. \nu_{TT} R_{TT}^{\ell} \sin^2 \theta \sin 2\phi \right] \\
 P_{t} \bar{\sigma} &=&  \nu_0 \left[ \nu_{LT} R_{LT}^{t} \sin\phi
 + \nu_{TT} R_{TT}^{t} \sin\theta \sin 2\phi \right]     \\
 P^\prime_n \bar{\sigma} &=& \nu_0 \left[
 \nu_{LT}^\prime R_{LT}^{\prime n} \sin\phi \right] \\
 P^\prime_{\ell} \bar{\sigma} &=&  \nu_0 \left[
             \nu_{LT}^\prime R_{LT}^{\prime {\ell}} \sin\theta \cos \phi + 
             \nu_{TT}^\prime R_{TT}^{\prime {\ell}}   \right] \\ 
 P^\prime_{t} \bar{\sigma} &=&  \nu_0 \left[
             \nu_{LT}^\prime R_{LT}^{\prime {t}} \cos \phi + 
             \nu_{TT}^\prime R_{TT}^{\prime {t}}  \sin\theta \right] 
\end{eqnarray}
\end{subequations}
where
$\bar{\sigma}$ is the virtual $\gamma N$ unpolarized cm cross section,
$A$ is the beam analyzing power, and 
$P$ and $P^\prime$ are helicity-independent and helicity-dependent
polarizations expressed in terms of longitudinal, normal, and transverse
basis vectors
$\hat{\ell} = \hat{p}_N$,
$\hat{n} \propto \hat{q} \times \hat{\ell}$, and 
$\hat{t} \propto \hat{n} \times \hat{\ell}$.
Here $\vec{q}$ is the momentum transfer in the lab and $\vec{p}_N$ is the
final nucleon momentum in the $\pi N$ cm frame.
The response functions, $R$, depend upon $W$, $Q^2$, and $\cos{\theta}$,
where $\theta$ is the pion angle relative to $\vec{q}$ in the cm frame; 
subscripts $L$ and $T$ represent longitudinal and transverse polarization 
states of the virtual photon while superscripts include the nucleon 
polarization component and/or a prime for beam polarization, as appropriate.
Rosenbluth separation of the combinations 
$\nu_T R_{L+T}=\nu_L R_L + \nu_T R_T$ 
and $\nu_T R^n_{L+T}=\nu_L  R_L^n + \nu_T R_T^n$ requires variation of the
beam energy, which was not performed in this experiment.
The cm phase space is given by $\nu_0 = k/q_0$, 
where $k$ and $q_0$ are the pion and equivalent real photon momenta,
while the kinematical factors
$\nu_T=1$, $\nu_{TT}=\epsilon$, 
$\nu^\prime_{TT}=\sqrt{1-\epsilon^2}$,
$\nu_L=\epsilon_S$, $\nu_{LT}=\sqrt{2\epsilon_S(1+\epsilon)}$, 
$\nu^\prime_{LT}=\sqrt{2\epsilon_S(1-\epsilon)}$
are elements of the virtual photon density matrix based upon the transverse
and scalar (longitudinal) polarizations,  
$\epsilon=(1+2\frac{{\bf q}^2}{Q^2}\tan^2{\frac{\theta_e}{2}})^{-1}$ and
$\epsilon_S=\epsilon Q^2/{\bf q}^2$.
Finally, $\phi$ is the angle between the scattering and reaction planes. 

The experiment was performed in Hall A of Jefferson Lab using standard 
equipment described in Ref. \cite{HallA-nim}.
A beam of $4531 \pm 1$ MeV electrons, with current ranging between 
about 40 and 110 $\mu$A, was rastered on a 15 cm LH$_2$ target.
The beam polarization, averaging 72\% for the first two running periods
and 65\% for the third, was measured nearly continuously using a
Compton polarimeter, with systematic uncertainties estimated to be
about 1\% \cite{Escoffier01}.

Scattered electrons and protons were detected in two high-resolution 
spectrometers, each equipped with a pair of vertical drift chambers for 
tracking and a pair of scintillation planes for triggering.
Protons were selected using the correlation between velocity and
energy deposition in plastic scintillators and pion production was
defined by cuts on missing mass and the correlation between missing energy 
and missing momentum.
The proton polarization was analyzed by a focal-plane polarimeter (FPP).
Detailed descriptions of the FPP and its calibration procedures can be
found in Refs. \cite{Roche-thesis,Punjabi05}.
The electron spectrometer remained fixed at $14.1^\circ$ with a 
central momentum of 3.66 GeV/$c$, while the proton spectrometer angle
and momentum were adjusted to cover the angular distribution.
Although the motion of the spectrometers was limited to
the horizontal plane, the boost from cm to lab 
focuses the reaction into a cone with an opening angle of only $13^\circ$ 
and provides enough out-of-plane acceptance to access all of the 
response functions, even those that vanish for coplanar kinematics. 
Cross sections were deduced by comparison with a Monte Carlo model
of the phase space and acceptance for each setting, 
including radiative corrections.
This model reproduces the observed distributions very well \cite{Chai-thesis}.

The nucleon polarization at the target in the cm frame was
deduced from the azimuthal distribution for scattering in the FPP using
the method of maximum likelihood.
The likelihood function takes the form
\begin{equation}
\label{eq:R_likelihood}
{\cal L} = \prod_{\rm events} \frac{1}{2\pi} \left(
1 + \xi + \eta \cdot R
\right)
\end{equation}
where $\xi$ represents the false (instrumental) asymmetry,
$R$ is a vector containing the response functions, and
$\eta$ is a vector of eventwise calculable coefficients that depend upon 
kinematical variables, differential cross section, 
beam polarization and helicity, FPP scattering angles and analyzing power, 
and spin transport matrix elements.
The system of equations derived from
$\partial \ln {\cal L} / \partial R_m = 0$
is solved using an iterative procedure.
The procedure was tested using pseudodata: 
a model was used to compute response functions for each accepted event, 
the predicted polarization was transported to the focal plane using
the same transport matrix as for the data analysis, and 
the azimuthal angle in the FPP was sampled according to its 
probability distribution.
The pseudodata were then analyzed in the same manner as real data.
We found that the model responses are recovered with fluctuations
consistent with the statistical uncertainties.
We also found that small deviations between acceptance-averaged and 
nominal $Q^2$ can be compensated using a dipole form factor.

\begin{figure*}
\centering
\includegraphics[angle=90,width=5.5in]{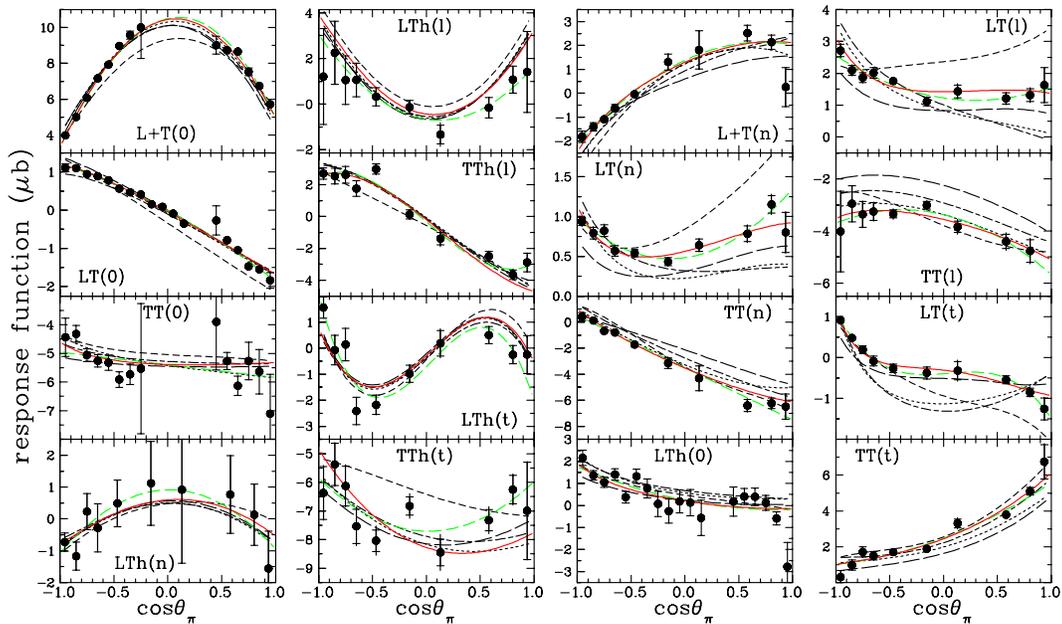}
\caption{(Color online) 
Data for response functions at $W = 1.23$ GeV and $Q^2 = 1.0$ (GeV/$c$)$^2$
are compared with recent models and with fits.
The labeling distinguishes L, T, LT, and TT contributions to the 
unpolarized (0) cross section and to transverse (t), normal (n), or 
longitudinal (l) components of recoil polarization with an h to indicate
helicity dependence, if any.
Linear combinations that cannot be resolved without Rosenbluth separation are
identified by L+T.
Black dash-dotted, dotted, short-dashed, and long-dashed curves represent the 
MAID2003, DMT, SAID and SL models, respectively.
The green mid-dashed curves show a Legendre fit while the solid red 
curves show a multipole fit.}
\label{fig:ndelta}
\end{figure*}

Systematic uncertainties due to acceptance normalization, FPP analyzing
power, beam polarization, elastic subtraction, false asymmetry, and 
spin rotation matrix elements were evaluated by comparing results from 
replays differing by a perturbation of the relevant parameter.
The propagation of systematic uncertainties for fitted Legendre
coefficients or multipole amplitudes was evaluated using fits to
those data sets.
Data for $W = 1.23$ GeV at $Q^2 = 1.0$ (GeV/$c$)$^2$ are shown in
Fig. \ref{fig:ndelta} for bin widths of $\Delta W = \pm 0.01$ GeV and 
$\Delta Q^2 = \pm 0.2$ (GeV/$c$)$^2$.
We show $R_{L+T}$, $R_{LT}$, and $R_{TT}$ extracted from the $\phi$ dependence 
of $\bar{\sigma}$ with error bars from fitting; the large error bars or 
missing bins for $\cos{\theta} \sim 0$ reflect inadequate $\phi$ coverage 
for this separation, but the phenomenological analyses use the actual
differential cross sections.
The bins of $\cos{\theta}$ for polarization were chosen to give 
approximately uniform statistics.
Inner error bars with endcaps show statistical uncertainties and outer error 
bars without endcaps include systematic uncertainties.
The systematic uncertainties in response functions and derived quantities 
are typically small compared with statistical or fitting uncertainties.

Figure \ref{fig:ndelta} also shows predictions from several recent models:  
MAID2003 \cite{Drechsel99,MAID2003}, DMT \cite{Kamalov01,DMT}, 
SAID \cite{Arndt03b,SAID}, and SL \cite{Sato01}.
Although the first three response functions in column 1 and the last in
column 3 have been observed before, the other 12 response functions have been
observed here for the first time.
The first two columns are determined by real parts of interference 
products and tend to be dominated by resonant amplitudes, while the last two
columns are determined by imaginary parts that are more sensitive to
nonresonant amplitudes.  
Thus, one finds relatively little variation among models for the first two 
columns and much larger variations for the last two,
but none provides a uniformly good fit, especially to imaginary responses.

Having factored out leading dependencies on $\sin\theta$, 
the response functions should be polynomials in $\cos{\theta}$ 
of relatively low order, especially if the assumption of $M_{1+}$ 
dominance is valid near the $\Delta$ resonance.
However, good fits over a range of $W$ require additional terms in
$R_{LT}$, $R_{TT}$, $R_{LT}^\prime$, $R_{L+T}^n$, and $R_{TT}^\ell$.
The green dashed curves fit coefficients of Legendre expansions to the data
for each response function independently, including terms beyond $M_{1+}$ 
dominance as needed; the extra terms have negligible effect upon the 
SMR and EMR values obtained using the traditional truncation formulas.
Our results for the Legendre analysis are compared with those of 
Joo {\it et al.}\ \cite{Joo02} for CLAS data in the top section of 
Table \ref{table:quad}.
These Legendre results for EMR overlap, but our result for SMR is 
more precise and significantly smaller.

The solid red curves in Fig. \ref{fig:ndelta} show a multipole analysis 
that varies the real and imaginary parts of all $s$-wave and $p$-wave 
amplitudes, except $\Im M_{1-}$, plus real parts of $2-$ multipoles. 
Higher partial waves were determined using a baseline model,
here based upon Born terms for pseudovector coupling.
We did not vary $\Im M_{1-}$ because all models considered predict that 
it is negligible for our $W$ range, yet experimentally it is strongly
correlated with $\Im S_{1-}$.
Note that we could not achieve acceptable multipole fits without 
varying the $s$-wave amplitudes with respect to baseline models and
we found that the imaginary part of $S_{0+}$ is especially important.
Small improvements for some of the responses can be obtained by varying
other $d$-wave amplitudes also, but the uncertainties in the quadrupole 
ratios increase because higher partial waves are not strongly constrained 
by these data and correlations between parameters become more severe. 
Fits starting from the MAID2003, DMT, or SL models are practically 
indistinguishable from those shown, but SAID is less suitable as a 
baseline model because some of its $\ell_\pi=2$ amplitudes are too large.
The insensitivity of quadrupole ratios to the choice of baseline model
is shown in Table \ref{table:quad}.
Therefore, the multipole analysis provides nearly model-independent
quadrupole ratios; we choose as final the results based upon the 
Born baseline to minimize residual theoretical bias.

Both Legendre and multipole analyses reproduce the data well but
the multipole analysis is more fundamental, 
employs fewer parameters (16 vs. 50), and
uses the data for all response functions simultaneously
while the more phenomenological Legendre analysis fits each response
function independently and ignores the relationships between Legendre 
coefficients required by expansions of those coefficients in terms of 
products of multipole amplitudes.
A more detailed paper is forthcoming that shows that neither assumption
of the traditional Legendre analysis ($sp$ truncation and $M_{1+}$ dominance)
is sufficiently accurate for data with the present levels of completeness 
and precision.
The relative error in the traditional Legendre analysis is particularly 
severe for EMR.

\begin{figure}
\centering
\includegraphics[width=2.75in]{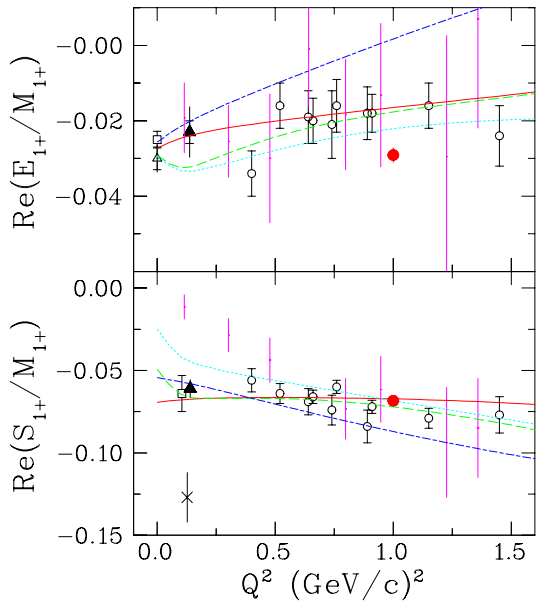}
\caption{(Color online) 
Red circles, present results from multipole analysis;
open squares, MAMI \protect{\cite{Beck00,Pospischil01}};
open triangle, LEGS \protect{\cite{Blanpied97}};
filled triangles, MIT \protect{\cite{Sparveris05}};
cross, ELSA \protect{\cite{Kalleicher97}};
open circles, CLAS \protect{\cite{Joo02}}.
Small horizontal displacements are used to reduce clutter.
Inner error bars with endcaps show statistical and systematic errors;
where available, outer error bars without endcaps include model error.
Red, green dashed, blue dash-dotted,  and cyan dotted curves represent 
MAID2003, DMT, SAID and SL, respectively.
Magenta bars show lattice QCD results \protect{\cite{Alexandrou05}}.}
\label{fig:quad}
\end{figure}

Recent data on quadrupole ratios for $Q^2 < 1.6$ (GeV/$c$)$^2$ are 
compared with representative models in Fig. \ref{fig:quad}.
Note that the MAID2003, DMT, and SL models included previous EMR and
SMR data in their parameter optimization.
The present result for EMR disagrees strongly with the SAID prediction
and is nearly identical to the data for $Q^2 = 0$, 
suggesting that EMR is nearly constant over this range.
Unlike the somewhat smaller CLAS results for EMR, our multipole result does 
not depend upon $sp$ truncation or $M_{1+}$ dominance.
Similarly, our SMR result is close to those for $Q^2<0.2$ (GeV/$c$)$^2$,
suggesting that SMR is nearly constant over this range also.
The stronger $Q^2$ dependence of lattice QCD calculations \cite{Alexandrou05}
may arise because the quenched approximation misses pionic contributions
that are expected to be important at low $Q^2$.

\begin{table}
\caption{Quadrupole ratios for $Q^2 = 1.0$ (GeV/$c$)$^2$.
The first section reports Legendre analyses and the second
multipole analyses based upon the specified baseline models.}
\label{table:quad}
\begin{ruledtabular}
\begin{tabular}{llll}
method/baseline & SMR, \% & EMR, \% & $\chi^2_\nu$ \\ \hline
Legendre   & $-6.11 \pm 0.11$ & $-1.92 \pm 0.14$ & 1.50 \\
Joo \etal \footnote{Weighted average of CLAS data for 
$Q^2 = 0.9$ (GeV/$c$)$^2$ from \cite{Joo02}.}
         & $-7.4  \pm 0.4 $ & $-1.8  \pm 0.4$ & \\ \hline
Born     & $-6.84 \pm 0.15$ & $-2.91 \pm 0.19$ & 1.65 \\
MAID2003 & $-6.90 \pm 0.15$ & $-2.79 \pm 0.19$ & 1.67 \\
DMT      & $-6.82 \pm 0.15$ & $-2.70 \pm 0.19$ & 1.67 \\
SL       & $-6.79 \pm 0.15$ & $-2.81 \pm 0.19$ & 1.64 \\
SAID     & $-7.38 \pm 0.15$ & $-2.53 \pm 0.20$ & 1.85 \\
\end{tabular}
\end{ruledtabular}
\end{table}

In summary, we have measured angular distributions of 14 separated 
response functions plus 2 Rosenbluth combinations for the 
$p(\vec{e},e^\prime \vec{p})\pi^0$ reaction at $Q^2= 1.0$ (GeV/$c$)$^2$ 
across the $\Delta$ resonance,
of which 12 have been obtained for the first time.
Dynamical models describe responses governed by real parts of interference
products relatively well, but differ both from each other and from the
data more strongly for imaginary parts that are more sensitive to
nonresonant mechanisms.
None of the theoretical models considered provides a uniformly good
description of the polarization data.
We performed a nearly model-independent multipole analysis and obtained 
$\text{SMR} = -(6.84 \pm 0.15)\%$ and $\text{EMR} = -(2.91 \pm 0.19)\%$.
The traditional Legendre analysis also fits the data well but gives
distinctly smaller quadrupole ratios, demonstrating that its assumptions
about the relative magnitudes and phases of multipoles are not 
sufficiently accurate.
A more detailed presentation of the fitted multipole amplitudes will
be given in a longer paper.

\begin{acknowledgments}
This work was supported by DOE contract No. DE-AC05-84ER40150 Modification
No. M175 under which the Southeastern Universities Research Association 
(SURA) operates the Thomas Jefferson National Accelerator Facility.  
We acknowledge additional grants from the U.S. DOE and NSF, the Canadian
NSERC, the Italian INFN, the French CNRS and CEA, and the Swedish VR.
\end{acknowledgments}


\begin{thebibliography}{10}

\bibitem{Isgur82}
N. Isgur, G. Karl, and R. Koniuk, Phys. Rev. {\bf D} {\bf {\bf 25}},  2394
  (1982).

\bibitem{Buchmann00}
A.~J. Buchmann and E.~M. Henley, Phys. Rev. {\bf C} {\bf {\bf 63}},  015202
  (2000).

\bibitem{Fiolhais96}
M. Fiolhais, G. Golli, and S. Sirca, Phys. Lett. {\bf {\bf B373}},  229
  (1996).

\bibitem{Kamalov99}
S.~S. Kamalov and S.~N. Yang, Phys. Rev. Lett. {\bf {\bf 83}},  4494  (1999).

\bibitem{Alexandrou05}
C. Alexandrou {\it et~al.}, Phys. Rev. Lett. {\bf {\bf 94}},  021601  (2005).

\bibitem{MAID2003}
D. Drechsel {\it et~al.}, \url{www.kph.uni-mainz.de/maid/maid2003}.

\bibitem{Kelly05d}
J.~J. Kelly {\it et~al.}, to be submitted to PRC (unpublished).

\bibitem{Warren98}
G.~A. Warren {\it et~al.}, Phys. Rev. {\bf C} {\bf {\bf 58}},  3722  (1998).

\bibitem{Pospischil01}
T. Pospischil {\it et~al.}, Eur. Phys. J. A {\bf {\bf 12}},  125  (2001).

\bibitem{Bartsch02}
P. Bartsch {\it et~al.}, Phys. Rev. Lett. {\bf {\bf 88}},  142001  (2002).

\bibitem{Joo03}
K. Joo {\it et~al.}, Phys. Rev. {\bf C} {\bf {\bf 68}},  032201  (2003).

\bibitem{Kunz03}
C. Kunz {\it et~al.}, Phys. Lett. {\bf {\bf B564}},  21  (2003).

\bibitem{Raskin89}
A.~S. Raskin and T.~W. Donnelly, Ann. Phys. (N.Y.) {\bf {\bf 191}},  78
  (1989).

\bibitem{HallA-nim}
J. Alcorn {\it et~al.}, Nucl. Instru. Meth. {\bf {\bf A522}},  294  (2004).

\bibitem{Escoffier01}
S. Escoffier, Ph.D. thesis, University of Paris, 2001.

\bibitem{Roche-thesis}
R.~E. Roch\'{e}, Ph.D. thesis, {Florida State University}, 2003.

\bibitem{Punjabi05}
V. Punjabi {\it et~al.}, Phys. Rev. {\bf C} {\bf {\bf 71}},  055202  (2005).

\bibitem{Chai-thesis}
Z. Chai, Ph.D. thesis, {Massachusetts Institute of Technology}, 2003.

\bibitem{Drechsel99}
D. Drechsel {\it et~al.}, Nucl. Phys. {\bf {\bf A645}}, 145 (1999).

\bibitem{DMT}
D. Drechsel {\it et~al.}, \url{www.kph.uni-mainz.de/maid/dmt}.

\bibitem{Kamalov01}
S.~S. Kamalov {\it et~al.}, Phys. Rev. {\bf C} {\bf {\bf 64}}, 032201(R) (2001).

\bibitem{SAID}
R.~A. Arndt {\it et~al.}, \url{gwdac.phys.gwu.edu}.

\bibitem{Arndt03b}
R.~A. Arndt {\it et~al.}, in {\em
  {Proceedings of the Workshop on the Physics of Excited Nucleons
  (NSTAR2002)}}, edited by S.~A. Dytman and E.~S. Swanson (World Scientific,
  Singapore, 2003).

\bibitem{Sato01}
T. Sato and T.-S.~H. Lee, Phys. Rev. {\bf C} {\bf {\bf 63}},  055201  (2001).

\bibitem{Joo02}
K. Joo {\it et~al.}, Phys. Rev. Lett. {\bf {\bf 88}},  122001  (2002).

\bibitem{Beck00}
R. Beck {\it et~al.}, Phys. Rev. {\bf C} {\bf {\bf 61}},  035204  (2000).

\bibitem{Blanpied97}
G. Blanpied {\it et~al.}, Phys. Rev. Lett. {\bf {\bf 79}},  4337  (1997).

\bibitem{Sparveris05}
N.~F. Sparveris {\it et~al.}, Phys. Rev. Lett. {\bf {\bf 94}},  022003  (2005).

\bibitem{Kalleicher97}
F. Kalleicher {\it et~al.}, Zeit. Phys. {\bf A} {\bf {\bf 359}},  201  (1997).

\end{thebibliography}

\end{document}